\documentstyle[fleqn,amssymb]{elsart}

\newcommand{\cB}{{\cal B}}

\newcommand{\cH}{{\cal H}}

\newcommand{\defeq}{:=}

\newcommand{\set}[1]{\{ #1 \}}

\def\bbz {{\mathbb Z}}

\newcommand{\abs}[1]{\left| #1 \right|}

\newcommand{\tr}[1]{{\rm tr}\, #1}

\def\idty{{\leavevmode{\rm 1\mkern -5.4mu I}}} 

\def\sg{\sigma}
\newcommand{\ul}[1]{\underline{#1}}

\newcommand{\ket}[1]{| #1 \rangle}

\newcommand{\braket}[2]{\langle #1 | #2 \rangle}

\begin{document}

\begin{frontmatter}

\title{Almost any quantum spin system\\
with short-range interactions\\
can support toric codes}

\author{Maxim Raginsky}

\address{Center for Photonic Communication and Computing\\
Department of Electrical and Computer Engineering\\
Northwestern University, Evanston, Illinois 60208-3118\\
E-mail address:  {\tt maxim@northwestern.edu}}

\begin{abstract}
Inspired by Kitaev's argument that physical error correction is possible in a system of interacting anyons, we demonstrate that such "self-correction" is fairly common in spin systems with classical Hamiltonians that admit the Peierls argument and where errors are modelled by quantum perturbations.


\end{abstract}

\end{frontmatter}

\section{Introduction}

Error correction is a key ingredient in any good recipe for quantum information processing (see \cite{nielchu} for an overview and further references).  Many ingenious schemes have been invented to that effect. A particularly interesting approach has been suggested by Kitaev and colleagues in a series of beautiful papers \cite{kit,bk,dklp}, namely the possibility of implementing quantum error correction {\em on the physical level}.

The idea is quite simple.  In \cite{kit}, Kitaev considers a specific Hamiltonian $H_\Lambda$ for a quantum spin-1/2 system on a finite lattice $\Lambda \subset \bbz^2$ with periodic boundary conditions (the spins are taken to reside on the bonds of $\Lambda$).  For any finite $\Lambda$, the ground state of $H_\Lambda$ is fourfold degenerate; information is stored in the corresponding four-dimensional eigenspace.  Addition of a small local perturbation of a certain kind modifies the Hamiltonian to $H_\Lambda(\epsilon)$, where $\epsilon$ is the perturbation strength.  The effect of the perturbation is to introduce an energy splitting between the degenerate ground-state levels of the unperturbed Hamiltonian.  Kitaev then argues that there exists a constant $\epsilon_0$ so that, at low temperatures and for all $\abs{\epsilon} \le \epsilon_0$, the energy splitting for sufficiently large $\Lambda$ is given by $\exp{(-c\sqrt{N_\Lambda})}$, where $N_\Lambda$ is the number of bonds in $\Lambda$. In other words, in the thermodynamic limit $\Lambda \uparrow \bbz^2$, the ground state is still fourfold degenerate, and any sufficiently weak perturbation is "washed out" by the system itself.

The four-body interactions comprising Kitaev's Hamiltonian were
originally considered by him in Ref.~\cite{kit2} as generators for a
family of stabilizer codes \cite{nielchu}, which he termed "toric
codes."  The remarkable feature of toric codes is the fact that,
despite their apparent nonoptimality (in the sense of \cite{cs}), they
require only local operations for their implementation and can correct
any number of errors (by making the lattice large enough). The bulk of
Kitaev's analysis of toric codes was concerned with their properties
as "conventional" quantum error-correcting codes \cite{kl} that
require active intervention through frequent measurements and other external processing. The issue of constructing "self-correcting" quantum spin systems on the basis of toric codes has been taken up again only very recently by Dennis, Kitaev, Landahl, and Preskill \cite{dklp}.  Their approach, however, is centered around topological features of toric codes and delves deep into such subjects as nonabelian gauge theory \cite{dklp,bgkp}.

On the other hand, the very idea of physical error correction is so tantalizing, both practically and conceptually, that one cannot help but wonder: how generic are phenomena of this kind?  In this Letter I show that a few results in statistical mechanics of quantum spin systems point towards the conclusion that physical error correction is fairly common in such systems, under quite reasonable conditions.  These preliminary findings are, in my opinion, suggestive and interesting enough, and might spur further research into this topic both in the quantum information community and in the statistical mechanics community.  A detailed analysis is forthcoming \cite{rag}.

\section{Laying out the ingredients}

First of all, let us agree on the ingredients necessary for analysis of a self-correcting quantum spin system.  Let $\Lambda \subset \bbz^\nu$ be a finite lattice.  We assume here that $\nu \ge 2$ is an integer.  Let $\cH_0$ be the $(2S+1)$-dimensional Hilbert space of a single particle of spin $S$. Spins are situated on lattice sites $\ell \in \Lambda$ (in Kitaev's construction, spins were located on lattice bonds).  In order to retain a superficial analogy with stabilizer codes, we will assume that the unperturbed Hamiltonian $H_\Lambda$ is {\em classical}, i.e., the interactions comprising it generate an abelian subalgebra of the algebra $\cB(\cH_\Lambda)$ of all linear operators on $\cH_\Lambda \defeq \bigotimes_{\ell \in \Lambda}\cH_\ell$, where $\cH_\ell$ is an isomorphic copy of $\cH_0$.  That is,
\begin{equation}
H_\Lambda = \sum_{M \subset \Lambda}\Phi_M,
\end{equation}
where each $\Phi_M$ is a Hermitian operator on $\cH_M \defeq \bigotimes_{\ell \in M}\cH_\ell$, and $[\Phi_M,\Phi_N] = 0$. We assume periodic boundary conditions (that is, the lattice $\Lambda$ is drawn on the torus $\bbz^\nu / \bbz$). We let $\set{\ket{\ul{\sg}}}$ be the orthonormal basis of $\cH_\Lambda$ in which $H_\Lambda$ is diagonal; the basis vectors are labelled by classical spin configurations, $\ul{\sg} = \set{\sigma(\ell)}_{\ell \in \Lambda}$ with $\sg(\ell) \in \set{-S,-S+1,\ldots,S-1,S}$.  We also assume that the smallest eigenvalue of $H_\Lambda$ is equal to zero, and that its geometric multiplicity is $m \ge 2$.  We denote the corresponding eigenspace by $\cH^g_\Lambda$.  

The effect of errors is modelled by introducing an off-diagonal perturbation term to the Hamiltonian:
\begin{equation}
H_\Lambda(\epsilon) \defeq H_\Lambda + \epsilon P,
\end{equation}
where $\epsilon$ is a positive constant and $P$ is a Hermitian operator whose exact form is, for the moment, left unspecified.  Addition of the $\epsilon P$ term will perturb the eigenvalues of $H_\Lambda$, resulting in energy splitting between orthogonal ground states of the original (unperturbed) Hamiltonian.  Consequently, we define
\begin{equation}
\Delta E_\Lambda(\epsilon) \defeq \max_{ \ket {\ul{\sg}}  \in \cH^g_\Lambda} \braket{\ul{\sg}}{H_\Lambda(\epsilon)|\ul{\sg}}.
\label{eq:esplit}
\end{equation}

The basic idea behind a self-correcting quantum spin system thus boils
down to the following.  Information is stored in the ground-state
eigenspace of the unperturbed Hamiltonian $H_\Lambda$.  The
multiplicity $m$ is, obviously, dictated by the desired storage
capacity:  when $m = 2^k$, our "ground-state memory cell" will hold
$k$ qubits.  Errors will cause some of the information to leak out
into excited states.  In order for error correction to take place, the
system should be able to recover its ground state from sufficiently
weak perturbations at sufficiently low temperature (the fact that we
have to work with low temperatures is clear since we deal with the
ground state).  That is, we hope that there exists a threshold value $\epsilon_0$ such that
\begin{equation}
\lim_{\Lambda \uparrow \bbz^\nu} \Delta E_\Lambda(\epsilon)=0,\qquad \epsilon \le \epsilon_0.
\label{eq:esplit_tl}
\end{equation}
This condition, however, is necessary but not sufficient for error
correction.  It may happen that the $m$-fold degeneracy of the ground
state does not survive in the thermodynamic limit, and we still lose
some storage capacity.  We therefore require that the ground state of
the perturbed Hamiltonian remains $m$-fold degenerate for all
$\epsilon \le \epsilon_0$ in the thermodynamic limit.  In the next section we elaborate further on these requirements for self-correction and show that they are quite easy to fulfill in a wide variety of quantum spin systems.  Before we proceed, we must remark that these are essentially "bare-bones" desiderata; more elaborate performance measures will also be investigated elsewhere \cite{rag}.

\section{Putting it together}

The main question is:  which restrictions ensue on the unperturbed Hamiltonian $H_\Lambda$ and on the perturbation $P$?  It turns out that this question can be answered using the same methods that are employed for constructing low-temperature phase diagrams for classical spin systems with quantum perturbations \cite{kt,bku,dff}.  Thus the Hamiltonian $H_\Lambda$ can be comprised by $n$-spin interactions (for fixed $n$) that satisfy the Peierls condition \cite{sla}:  the energy "cost" of a local perturbation $\omega'$ of a translationally invariant ground state $\omega$ is of the order of the surface area of the region that encloses the part of the lattice on which $\omega$ and $\omega'$ differ.  Additionally, the unperturbed Hamiltonian is assumed to have a spectral gap $g > 0$ (i.e., its first nonzero eigenvalue $\ge g$).  Admissible perturbations are formed by sums of translates of an arbitrary Hermitian operator $P_0$, whose support (the set of sites on which the action of $P_0$ is nontrivial) is finite and encloses the origin of the lattice.  Thus
\begin{equation}
P = \sum_{\ell \in \Lambda} P_\ell,
\end{equation}
where $P_\ell = \gamma(\ell) P_0$ with $\gamma(\ell)$ being the automorphism induced by the translation of the lattice $\Lambda$ that maps the origin $0$ to the site $\ell$ and respects the periodic boundary conditions.  Also, both the unperturbed and the perturbed Hamiltonians are assumed to be invariant under unitary transformations induced by a symmetry group acting transitively on the set $\set{\ket{\ul{\sg}}} \cap \cH^g_\Lambda$.

Assuming these conditions are satisfied, we invoke the following
theorem \cite{kt}:  there exists a constant $\epsilon_0$ such that,
for all $\epsilon \le \epsilon_0$, the perturbed spin system has $m$
translationally invariant ground states in the thermodynamic limit.
Furthermore, if the $m$ translationally invariant ground states of the
unperturbed Hamiltonian are invariant under some additional symmetries, these invariance properties carry over to the ground states of the perturbed system.  The threshold value $\epsilon_0$ of the parameter $\epsilon$ is determined by developing a low-temperature expansion \cite{gin,kt} of the perturbed partition function using a modified Lie-Trotter product formula,
\begin{equation}
\tr{e^{-\beta H_\Lambda(\epsilon)}} = \lim_{N \rightarrow \infty} \tr {\left \{ \left[ \left(\idty - \frac{\epsilon P}{N} \right) e^{-(1/N)H_\Lambda} \right]^{N\beta} \right\} }.
\end{equation}
The trace is expanded in the basis $\set{ \ket{\ul{\sg}}}$, thus
allowing for combinatorial analysis on a "space-time" grid, where the space axis is labelled by lattice sites $\ell$ and the time axis is labelled by values $0,(1/N)\beta,(2/N)\beta,\ldots,\beta$. The perturbation theory is controlled by a suitable coarse-graining of the time axis, which then allows to determine the threshold value $\epsilon_0$ that will render the contributions of the perturbation terms $P_\ell$ sufficiently small. (See \cite{gin,kt} for details; we only remark here that translation invariance requirements can be lifted, the perturbation theory still goes through \cite{kt}.)  Similar space-time analysis of the error rate has been described heuristically by Dennis et al. \cite{dklp}.

Another important issue is the following:  while the infinite-volume
ground state of the perturbed system retains the degeneracy of the
original (unperturbed) ground state, the degeneracy may be lost in the
case of finite lattice size.  This phenomenon is referred to as
"obscured symmetry breaking" \cite{kt2}, whereby low-lying eigenstates
of the finite-system Hamiltonian converge to additional ground states in the thermodynamic limit.  In this case we will have, for any finite $\Lambda$, 
\begin{equation}
\Delta E_\Lambda(\epsilon) > 0.
\end{equation}
It is therefore important to obtain an estimate of the convergence rate in (\ref{eq:esplit_tl}); namely, given some $\delta > 0$, find $N_0$ such that
\begin{equation}
\Delta E_\Lambda(\epsilon) < \delta,\qquad \abs{\Lambda} \ge N_0.
\end{equation}
Knowing the convergence rate allows us to appraise the resources needed to implement error correction with the desired accuracy $\delta$.  In this respect, an estimate of the form
\begin{equation}
\Delta E_\Lambda(\epsilon) = e^{-c \abs{\Lambda}},\qquad \epsilon \le \epsilon_0,
\label{eq:expcon}
\end{equation}
where the constant $c$ depends on $\epsilon$, would be ideal --- an exponential gain in error-correction accuracy could then be achieved with polynomial resources.  This exponential convergence rate is, in fact, one of the most attractive features of Kitaev's construction in Ref.~\cite{kit}.  

On the other hand, the rate at which $\Delta E_\Lambda(\epsilon)$ converges to zero is determined by the unperturbed Hamiltonian $H_\Lambda$, the perturbation $P$, and the perturbation strength $\epsilon$.  It is therefore important to know what we can expect in a generic setting. Obviously, the exponential behavior, as in Eq.~(\ref{eq:expcon}), is optimal, but it may as well turn out that the particular implementation (e.g., with a different Hamiltonian) does not allow for it. We can, however, hope for a slower (but still quite decent) convergence rate.  According to a theorem of Horsch and von der Linden \cite{hl}, certain quantum spin systems possess low-lying eigenstates of the finite-lattice Hamiltonian with
\begin{equation}
\Delta E_\Lambda(\epsilon) = c/\abs{\Lambda}.
\end{equation}
The conditions for this to hold are the following.  There has to exist an {\em order observable} $O_\Lambda$ of the form
\begin{equation}
O_\Lambda = \sum_{\ell \in \Lambda} O_\ell,
\end{equation}
where each $O_\ell$ is a Hermitian operator such that $[O_\ell,
O_{\ell '}] = 0$.  Furthermore, for any interaction term $\Phi_{M
\subset \Lambda}$ in the perturbed Hamiltonian $H_\Lambda(\epsilon)$
(these also include the perturbation terms), we will have $[\Phi_M,O_\ell]
= 0$ unless $\ell \in M$.  The operators $\Phi_M$ and $O_\ell$ are
required to be uniformly bounded (in $M$ and $\ell$ respectively), and
the cardinality of the support set $M$ must not exceed some fixed
constant $C$ (the latter condition has also to be fulfilled for the
perturbation theory described above to converge). Finally, if
$\ket{\varphi}$ is an eigenstate of $H_\Lambda(\epsilon)$, then we
must have $\braket{\varphi}{O_\Lambda | \varphi} = 0$, but
$\braket{\varphi}{O^2_\Lambda | \varphi} \ge \zeta \abs{\Lambda}^2$
(here the constant $\zeta$ depends on $O_\ell$).  The latter
conditions are taken as manifestations of "obscured symmetry
breaking'' \cite{kt2}. Examples of systems for which the Horsch-von der Linden theorem holds include \cite{kt2} the Ising model in the transverse magnetic field or the Heisenberg antiferromagnet with a N\'eel order.

\section{Conclusion}

Where does it all take us?  It appears, from the discussion in the
previous section, that any quantum spin system, whose Hamiltonian is
formed by mutually commuting $n$-body interactions that satisfy the
Peierls condition, can recover from sufficiently weak quantum (i.e.,
off-diagonal) perturbations at low temperatures.  The admissible
perturbations can be either finite-range \cite{kt}, or exponentially
decaying \cite{bku}.  Under these (quite general) conditions, it
follows from rigorous perturbation theory for quantum spin systems
that there exists a critical perturbation strength $\epsilon_0$, such
that for all $\epsilon < \epsilon_0$ the degeneracy and symmetry properties of the ground state of the original (unperturbed) system survive in the thermodynamic limit.  Furthermore, even if ground-state degeneracy is removed by perturbation of the finite-size system, the effect of the error (perturbation) is effectively "washed out" in the thermodynamic limit, as the low-lying excited states of the perturbed system converge to additional ground states.

However, the systems we have considered in this Letter are assumed to have classical Hamiltonians and discrete symmetries.  What about truly quantum Hamiltonians and continuous symmetry (e.g., the quantum Heisenberg model)?  The situation here is not so easy.  For instance, it is apparent from our discussion that, in order to be self-correcting, the perturbed system must exhibit an order-disorder transition as the parameter $\epsilon$ is tuned: in the "ordered phase," error correction is possible; in the "disordered phase," occurrence of errors results in irrevocable loss of information.  (This has already been noted by Dennis et al. \cite{dklp}.)  Since we require the ground state of the perturbed Hamiltonian to exhibit the same degeneracy as the corresponding state of the unperturbed Hamiltonian, it makes sense to talk about spontaneously broken symmetry in the ordered phase (i.e., when $\epsilon < \epsilon_0$).  However, according to the so-called Goldstone theorem \cite{lfw}, symmetry cannot be broken in a system with continuous symmetry and a gap.  It would certainly be worthwile to explore physical error correction in systems with continuous symmetries as well, but the models in which it can work will not be as easy to find.

\section*{Acknowledgments}

The author has benefitted from a few discussions with V. Moroz. This work was supported by the U.S. Army Research Office through MURI grant DAAD19-00-1-0177.

\end{document}